# Solving Assembly Line Balancing Problems by Combining IP and CP

Alexander Bockmayr and Nicolai Pisaruk[*]

Université Henri Poincaré, LORIA
B.P. 239, F-54506 Vandœuvre-lès-Nancy, France
{bockmayr|pisaruk}@loria.fr

**Abstract.** Assembly line balancing problems consist in partitioning the work necessary to assemble a number of products among different stations of an assembly line. We present a hybrid approach for solving such problems, which combines constraint programming and integer programming.

## 1 Introduction

Assembly lines are special flow-line production systems typical for the industrial production of high quantity standardized commodities. An assembly line consists of a number of work stations arranged along a conveyor belt. The work pieces are consecutively launched down the conveyor belt and are moved from one station to the next. At each station, one or several tasks necessary to manufacture the product are performed. The problem of partitioning the various tasks among the stations with respect to some objective function is called the assembly line balancing problem (ALBP) [16].

Various classes of assembly line balancing problems have been studied in the literature. We will consider here so-called *simple assembly line balancing problems* SALBP of the following form: Let $M = \{1, \ldots, m\}$ be the set of stations and $N = \{1, \ldots, n\}$ be the set of tasks. We denote by $t_j$ the time required for task $j$ and by $S_j \subseteq M$ the set of stations able to fulfill task $j$. There is a precedence relation on the tasks that will be represented by a graph $G = (N, E)$, where $(j_1, j_2) \in E$ means that task $j_1$ is an immediate predecessor of task $j_2$. By $CT_i$ we denote the total time available for executing the tasks that have been assigned to station $i$. The capacities $CT_i$ may vary from one station to another due to, e.g., different numbers of operators.

The constraint satisfaction problem in assembly line balancing consists in assigning tasks to stations such that the total running time of the tasks assigned to some station does not exceed its capacity, and such that the precedence relations between the tasks are satisfied. The objective function is to minimize the number of stations necessary to fulfill all tasks. Including bin packing as a special case, the assembly line balancing problem is NP-hard.

[*] This work was partially supported by the European Commission, Growth Programme, Research Project LISCOS – Large Scale Integrated Supply Chain Optimisation Software, Contract No. G1RD-CT-1999-00034

The aim of this paper is to present a hybrid solver for assembly line balancing problems, which combines constraint programming (CP) and integer programming (IP). The integration of integer programming and constraint programming has been an important research topic during the last years, see e.g. [15,4,9,10,7,14]. The contribution of this paper is twofold: we develop a branch-and-cut solver for SALBP and show how it can cooperate with a CP solver in order to prune the search tree.

The organization of the paper is as follows. We start in Sect. 2 with an integer programming model of the simple assembly line balancing problem. Sect. 3 describes the cutting planes that are used in the branch-and-cut solver on the IP side. Sect. 4 introduces a CP model of SALBP and Sect. 5 describes the cooperation between the IP and the CP solver. Finally, Sect. 6 contains a number of empirical results illustrating the benefits of the approach.

## 2 Integer Programming Model

Let $T_i = \{j : i \in S_j\}$ be the set of tasks which can be carried out by station $i$, and let $A = \cup_{j \in N}(S_j \times \{j\})$. We define the following decision variables

$$x_{ij} = \begin{cases} 1, & \text{if task } j \text{ is assigned to station } i, \\ 0, & \text{otherwise.} \end{cases}$$

### 2.1 Constraints

Feasible solutions $x = [x_{ij}]$ of the assembly line balancing problem have to satisfy the following constraints:

$$\sum_{i \in S_j} x_{ij} = 1, \quad j \in N, \tag{1}$$

$$\sum_{j \in T_i} t_j x_{ij} \leq CT_i, \quad i \in M, \tag{2}$$

$$\sum_{i \in S_{j_1},\, i \leq k} x_{ij_1} - \sum_{i \in S_{j_2},\, i \leq k} x_{ij_2} \geq 0, \quad k = 1, \ldots, m,\ (j_1, j_2) \in E, \tag{3}$$

$$x_{ij} \in \{0, 1\}, \quad (i, j) \in A. \tag{4}$$

SOS (Special Ordered Set) constraints (1) ensure that each task is assigned to exactly one workstation. Knapsack constraints (2) guarantee that the total running time of the tasks assigned to some station does not exceed its capacity. The constraints (3) correspond to the precedences given in the graph $G$. They express that if $(j_1, j_2) \in E$ and task $j_2$ is assigned to station $k$, then task $j_1$ must be assigned to one of the stations $1, \ldots, k$.

### 2.2 Objective function

Our objective is to minimize the number of stations necessary to perform all the tasks. Assuming w.l.o.g. that at most $n$ stations are available, we define costs

$c_{ij}$ for assigning task $j$ to station $i$ that satisfy the following condition:

$$c_{ij} = c_i \text{ for all } j \in T_i; \quad nc_i \leq c_{i+1}, \; i = 1, \ldots, m - 1. \tag{5}$$

This ensures that lower numbered stations will be used first. Now the objective function can be written as follows:

$$\min \sum_{j \in N} \sum_{i \in S_j} c_i x_{ij} \,. \tag{6}$$

## 3  Valid Inequalities and Cut Generation

In order to solve the IP model presented above, we will use a *branch-and-cut* approach [18]. In this section, we present the different classes of inequalities that will be used in our branch-and-cut algorithm. We start by defining the SALB *polytope*

$$P_{\text{SALB}} \stackrel{\text{def}}{=} \text{conv}(\{x \in \{0,1\}^A : x \text{ satisfies } (2),(3),(7)\}).$$

Here conv$(S)$ denotes the convex hull of a set of points $S \subset \mathbb{R}^n$. We relaxed the SOS constraints (1) to the inequalities

$$\sum_{i \in S_j} x_{ij} \leq 1, \quad j \in N. \tag{7}$$

Since the polytope $P_{\text{SALB}}$ is contained in the *multiple knapsack polytope* (MK), defined as the convex hull of the set of points $x \in \{0,1\}^A$ satisfying (7) and (2), all inequalities valid for the MK polytope are also valid for the SALB polytope. Furthermore, the multiple knapsack problem is a special type of the *generalized assignment problem* (GAP); therefore, the inequalities valid for the GAP polytope are also valid for the SALB polytope. The known classes of inequalities for the GAP and MK polytopes are based on the notion of cover, which we briefly recall in the next section.

### 3.1  Lifted Cover and (1,d)-Configuration Inequalities

For $a \in \mathbb{R}^n_{>0}$ and $b \in \mathbb{R}_{>0}$, let

$$P(a,b) \stackrel{\text{def}}{=} \text{conv}\left\{x \in \{0,1\}^n : \sum_{j=1}^n a_j x_j \leq b\right\}$$

denote the *knapsack polytope*. A subset $C \subseteq N$ is called a *cover* if $\sum_{j \in C} a_j > b$; a cover $C$ is *minimal* if $C \setminus \{s\}$ is not a cover for all $s \in C$. For a cover $C$, the *cover inequality*

$$\sum_{j \in C} x_j \leq |C| - 1 \tag{8}$$

is valid for $P(a,b)$; moreover, if $C$ is a minimal cover, then (8) defines a facet of

$$P^C(a,b) \stackrel{\text{def}}{=} P(a,b) \cap \{x \in \mathbb{R}^n : x_j = 0,\ j \in N \setminus C\}.$$

A pair $(H, z)$ is called a $(1, d)$-*configuration* [12] if $H \subset N$, $z \in N \setminus H$, and $2 \leq d \leq |H|$ are such that

- $\sum_{j \in H} a_j \leq b$;
- $H' \cup \{z\}$ is a minimal cover for every $H' \subseteq H$ with $|H'| = d$.

If $(H, z)$ is a $(1, d)$-configuration, the inequality

$$\sum_{j \in H} x_j + (|H| - d + 1)x_z \leq |H|$$

defines a facet of $P^{H \cup \{z\}}(a, b)$.

Lifting of inequalities is a key issue in branch-and-cut. It allows one to strengthen an inequality by calculating non-zero coefficients for variables that initially are not present. For a formal definition and general results about lifting see [18]. The lifted cover and (1,d)-configuration inequalities are automatically generated by the solver we have used. For the precedence-constrained knapsack polytope, a straightforward generalization of minimal covers, the so-called minimal induced covers, has been investigated, see e.g. [5,13,17]. By analogy, we introduce the *induced cover inequalities* for the SALB polytope.

Two tasks $j_1, j_2 \in N$ are called *incomparable* if both $(j_1, j_2) \notin E$ and $(j_2, j_1) \notin E$. A set $W \subseteq N$ is called *incomparable* if the elements in $W$ are pairwise incomparable.

To each station $k \in M$ corresponds a knapsack given by the inequality $\sum_{j \in T_k} t_j x_{kj} \leq CT_k$. We say that $C \subseteq T_k$ is a *minimal induced cover* (MIC) [5] if

- $C$ is incomparable,
- $\sum_{j \in C^{\leq}} t_j x_{kj} > CT_k$,
- $\sum_{j \in C^{\leq} \setminus \{s\}} t_j x_{kj} \leq CT_k$ for all $s \in C$.

Here $C^{\leq} \stackrel{\text{def}}{=} \{j \in N : j \leq j_1 \text{ for some } j_1 \in C\}$.

Let $C \subseteq N$ be a MIC for knapsack $k$. Then the inequality

$$\sum_{j \in C} x_{kj} - \sum_{j \in C^{\leq} \setminus C} \sum_{i \in S_j : i < k} x_{ij} \leq |C| - 1.$$

is valid for the SALB polytope.

### 3.2 Cycle Inequalities

The class of cycle inequalities was introduced in [8] for the GAP polytope. Since we are not aware of a separation algorithm for this class, we describe here a separation heuristic for the subclass of cycle inequalities with cycles of length 4.

Given two tasks $u$ and $v$; assume that $t_u \leq t_v$. Let $C_k$ and $C_l$ be covers of the knapsacks $k$ and $l$ resp. such that: a) $u, v \in C_k$, b) $u \in C_l$, $v \notin C_l$, c) $C_k \cap C_l = \{u\}$. Then

$$\sum_{j \in C_k} x_{kj} + \sum_{j \in C_l \cup \{v\}} x_{lj} \leq |C_k| + |C_l| - 3 \tag{9}$$

is a valid inequality for the SALB polytope [8].

**4-Cycle Heuristic**

- Choose two knapsacks $k, l \in M$, and two items $u, v \in T_k \cap T_l$ such that, for the point $x = [x_{ij}]$ to be separated, all four values $x_{ku}$, $x_{lu}$, $x_{kv}$, $x_{lv}$ are greater than zero. Assume that $t_u \leq t_v$.
- Compute a minimal cover $C'_k$ of the knapsack $\sum_{j \in T_k \setminus \{u,v\}} t_j x_{lj} \leq CT_l - t_u - t_v$ by solving

$$\min_{C \subseteq T_k \setminus \{u,v\}} \sum_{j \in C} (1 - x_{kj}) \;:\; \sum_{j \in C} t_j \geq CT_k - t_u - t_v + 1. \tag{10}$$

  Set $C_k = C'_k \cup \{u, v\}$.
- Compute a minimal cover $C'_l$ of the knapsack $\sum_{j \in T_l \setminus C_k} t_j x_{lj} \leq CT_k - t_u$ by solving

$$\min_{C \subseteq T_l \setminus C_k} \sum_{j \in C} (1 - x_{lj}) \;:\; \sum_{j \in C} t_j \geq CT_k - t_u + 1. \tag{11}$$

  Set $C_l = C'_l \cup \{u\}$.
- If both covers $C_k$ and $C_l$ exist, lift inequality (9) using the procedure described in section 3.4; otherwise, return failure.

### 3.3 Extended Cover and Heterogeneous Two-Cover Inequalities

Let $C$ be a cover with respect to some knapsack $k \in M$, and $D \subseteq N \setminus C$ be such that $D \cup \{i\}$ is a cover of knapsack $l \in M \setminus \{k\}$, for all $i \in C$. Then the *extended cover inequality*

$$\sum_{j \in C} (x_{kj} + x_{lj}) + \sum_{j \in D} x_{lj} \leq |C| + |D| - 1$$

is valid for the multiple knapsack polytope [6].

Consider two knapsacks $k, l \in M$, $k \neq l$. Assume that $C \subseteq T_k$ is a cover with respect to knapsack $k$, and let $D$ be a subset of $T_l \setminus C$ such that for all $D' \subseteq D$ and $C' \subseteq C$ with $|C'| = |D'|$, the set $(C \setminus C') \cup D'$ is a cover for knapsack $l$. Then the *heterogeneous two-cover inequality*

$$\sum_{j \in C} x_{kj} + \sum_{j \in C \cup D} (|C| - 1)\, x_{lj} \leq |C|\,(|C| - 1)$$

is valid for the multiple knapsack polytope [6]. For separation of extended cover and heterogeneous two-cover inequalities, we use the heuristics described in [6].

### 3.4 A General Lifting Procedure

Let $\bar{A} \subset A$ and

$$\sum_{(i,j) \in \bar{A}} \alpha_{ij} x_{ij} \leq \beta \tag{12}$$

be a valid inequality for $P_{\text{SALB}}$. For $(i_0, j_0) \in A \setminus \bar{A}$, the inequality

$$\alpha_{i_0 j_0} x_{i_0 j_0} + \sum_{(i,j) \in \bar{A}} \alpha_{ij} x_{ij} \leq \beta \tag{13}$$

if valid for $P_{\text{SALB}}$ if the coefficient $\alpha_{i_0 j_0}$ is computed by the following procedure:

**Lifting Heuristic**

- Let $A^* = \bar{A}$. For $(i,j) \in \bar{A}$, if $j = j_0$, or $i_0 < i$ and $(j, j_0) \in E$, or $i_0 > i$ and $(j_0, j) \in E$, exclude $(i, j)$ from $A^*$.
- For each $i \in M$,
  - generate a permutation $\pi^i = (\pi_1^i, \ldots, \pi_{n_i}^i)$ of the elements of $T_i$, $n_i = |T_i|$;
  - for $s = 2, \ldots, n_i$, compute

$$v_{i,s}^{\pi} \stackrel{\text{def}}{=} \max \left\{ \sum_{j=1}^{s} y_j : y \in \{0,1\}^s, \sum_{j=1}^{s} t_{\pi_j^i} y_j \leq CT_i \right\}. \tag{14}$$

- Solve the optimization problem

$$\begin{aligned}
\gamma = \max & \sum_{(i,j) \in A^*} \alpha_{ij} z_{ij} \\
& \sum_{i: (i,j) \in A^*} z_{ij} \leq 1, \quad j \in N, \\
& \sum_{\substack{1 \leq j \leq s, \\ (i, \pi_j^i) \in A^*}} z_{i\pi_j^i} \leq v_{i,s}^{\pi}, \quad s = 2, \ldots, n_i; \ i \in M, \\
& z_{ij} \in \{0,1\}, \ (i,j) \in A^*.
\end{aligned} \tag{15}$$

and set $\alpha_{i_0 j_0} = \beta - \gamma$.

Note that both optimization problems (14) and (15) can be solved in polynomial time. Problem (14) is a special type of knapsack problem and can be solved efficiently by dynamic programming. Problem (15) is a special case of the weighted matroid intersection problem [11]. In fact, it can also be reduced to the minimum cost maximum flow problem, see e.g. [2].

## 4 Constraint Programming Model

Our CP formulation of the SALB problem is based on the `cumulative` constraint [1]. The basic version of `cumulative` can be defined as follows.

There are $n$ tasks; task $j$ is characterized by three parameters, which can be either domain variables or values: the starting time $start_j$, the duration $dur_j$, and the amount $res_j$ of some resource consumed by the task. We are also given the completion time $e$ for all the tasks, and the upper bound $v$ on the resource consumption; $e$ and $v$ again are domain variables or values. The global constraint

$$\texttt{cumulative}([[start_1, dur_1, res_1], \ldots, [start_n, dur_n, res_n]], v, e)$$

is satisfied if the following conditions hold:

$$\sum_{\substack{1 \leq j \leq n:\\ start_j \leq t < start_j + dur_j}} res_j \leq v, \quad t = 1, \ldots, e,$$

$$\max_{1 \leq j \leq n} (start_j + dur_j) \leq e.$$

In addition to the $n$ given tasks, we introduce $m$ artificial tasks numbered $n+1, \ldots, n+m \stackrel{\text{def}}{=} \overline{n}$; for $CT^{\max} = \max_{i \in M} CT_i$, let $S_{n+i} = \{i\}$, $t_{n+i} = CT^{\max} - CT_i$, $i \in M$. Each task $j \in \{1, \ldots, n+m\}$ is associated with a triple $(start_j, dur_j, res_j)$ of domain variables, where

- $start_j = i$ if task $j$ is assigned to station $i$,
- $dur_j = 1$,
- $res_j = t_j$.

Now the CP model can be stated as follows:

Variables: $\begin{cases} start_j \in S_j, & j = 1, \ldots, \overline{n}, \\ dur_j \in \{1\}, & j = 1, \ldots, \overline{n}, \\ res_j \in \{t_j\}, & j = 1, \ldots, \overline{n}. \end{cases}$

Constraints: $\begin{cases} \texttt{cumulative}([[start_1, dur_1, res_1], \ldots, [start_{\overline{n}}, dur_{\overline{n}}, res_{\overline{n}}]], \\ \quad CT^{\max}, m+1); \\ start_{j_1} \leq start_{j_2}, \quad (j_1, j_2) \in E. \end{cases}$

## 5 Combining IP and CP

### 5.1 Reducing the Problem Size

The size of the IP and CP model described before can be reduced as follows.

In a first step, we build the CP model and do propagation, but without labeling. Let $S_j$ denotes the domain of $start_j$ after propagation, and let $T_i = \{j : i \in S_j\}$. The IP model is obtained by imposing the constraints (1)–(4) for these reduced sets $S_j$ and $T_i$.

In a second step, we do propagation by combining both solvers. Iteratively, for each task $j \in N$, we perform the following operations:

- Minimize $\sum_{i \in S_j} i \cdot x_{ij}$ subject to the constraints (1)–(4) using cut generation, but without branching. Let $\gamma_1$ denote the objective value returned by the solver. For $i \in S_j$ and $i < \lfloor \gamma_1 \rfloor$, set $x_{ij} = 0$ and remove $i$ from the domain of $start_j$.

- Maximize $\sum_{i \in S_j} i \cdot x_{ij}$ under the constraints (1)–(4) using cut generation, but without branching. Let $\gamma_2$ denote the objective value returned by the solver. For $i \in S_j$ and $i > \lceil \gamma_2 \rceil$, set $x_{ij} = 0$ and remove $i$ from the domain of $start_j$.
- Initiate propagation for the CP problem. For $q \in N$, if $i$ is not in the domain of $start_q$, set $x_{iq} = 0$.

### 5.2 Propagation and "Rounding Off" LP solutions

After reducing the problem size, we can use both solvers, IP or CP, to continue. If we decide to use the CP solver, there is no further interaction with the IP solver.

If we use the IP solver, we propose the following cooperation between IP and CP. Assume that $S_j$, for $j \in N$, is the domain of $start_j$ after preprocessing. For these sets $S_j$, consider the IP problem (6),(1)–(4).

**Propagation for subproblems.** The IP procedure starts processing a node of the branch-and-cut tree by calling the CP solver in the following way. For the LP subproblem at this node, let $d_{ij}^-, d_{ij}^+ \in \{0, 1\}$ denote the current lower and upper bounds of the variables $x_{ij}$. For $j \in N$, set $S_j = \{i : d_{ij}^+ = 1\}$. For these sets $S_j$, build the CP problem and do propagation. At the end, let $d_{ij}^+ = 0$ for those $i \in S_j$ that are not in the domain of $start_j$.

**Rounding of LP solutions.** Let $x = [x_{ij}]$ be a solution of some LP subproblem. For $j \in N$, define $S_j = \{i : x_{ij} > 0\}$, build the CP model, and run labeling for a limited amount of time, trying to find a feasible solution.

## 6 Empirical Results

This paper describes ongoing work. Therefore, we can give only preliminary empirical results. These are based on a selection of benchmarks from the data sets in [16]. For these examples, the cycle time $CT$ is the same for all stations, i.e., $CT = CT_i, i \in M$. The computational experiments were done on a Pentium III 600 MHz using the CHIP C Library [3] and the branch-and-cut code developed by the second author.

We compare two versions of the problem reduction heuristics described in Sect. 5.1. The first version generates cuts only when computing an initial feasible LP solution. All other LP problems are solved without generating cuts (except standard lifted cover and $(1, d)$-configuration cuts, which are automatically produced by the IP solver). The second version generates all the different cuts for all LP problems.

We measure the quality of a problem reduction strategy by the sum of the cardinalities of the sets $S_j$ upon termination of the procedure. These values are given in the last three columns of Table 1 for CP alone, CP and LP with standard cuts, and finally for CP and LP with all cuts.

| Instance name | # tasks n | Cycle time $CT$ | # stations m | Initial size | After CP | After LP with standard cuts | all cuts |
|---|---|---|---|---|---|---|---|
| Sawyer30 | 30 | 47 | 7 | 210 | 112 | 109 | 33 |
| Sawyer30 | 30 | 28 | 12 | 360 | 187 | 187 | 179 |
| Gunther | 35 | 54 | 9 | 315 | 129 | 105 | 105 |
| Gunther | 35 | 44 | 12 | 420 | 182 | 182 | 176 |
| Lutz3 | 89 | 118 | 14 | 1246 | 285 | 262 | 195 |
| Lutz3 | 89 | 74 | 23 | 2047 | 386 | 386 | 318 |
| Warnecke | 58 | 155 | 10 | 580 | 248 | 241 | 212 |
| Warnecke | 58 | 73 | 22 | 1276 | 575 | 575 | 575 |
| Tonge70 | 70 | 251 | 14 | 980 | 394 | 394 | 394 |

**Table 1.** Reducing problem size

Next we give in Table 2 the total running times for the hybrid solver described in Sect. 5.2 including problem reduction by CP propagation.

| Instance name | CP prop. time | Cut gener. time | Total sol. time |
|---|---|---|---|
| Sawyer30 (47-7) | 1.261 | 2.535 | 5.809 |
| Sawyer30 (28-12) | 0.260 | 0.901 | 2.824 |
| Gunther (54-9) | 0.031 | 0.230 | 0.401 |
| Gunther (44-12) | 0.191 | 0.941 | 1.712 |
| Lutz3 (118-14) | 0.821 | 14.151 | 16.183 |
| Lutz3 (74-23) | 1.895 | 42.019 | 47.849 |
| Warnecke (155-10) | 5.597 | 17.367 | 38.345 |
| Warnecke (73-22) | 2.765 | 29.007 | 1:36.499 |
| Tonge70 (251-14) | 1:27.120 | 10:22.7 | 13:49.330 |

**Table 2.** Running time in min:sec

Note that neither the CP nor the IP solver alone are able to solve, for example, problem Warnecke (73-22) in less than one hour of running time.

## 7 Conclusion and Further Research

Current computational experience does not allow us to tell whether CP and IP, alone or in cooperation, are able to solve to optimality the unsolved instances of the simple assembly line balancing problem in the data sets from [16]. Further research and implementation work is necessary to answer this question, in particular when the precedence relations are not tight.

While this is an interesting theoretical question, the main impact of this work is practical. The hybrid IP/CP solver described in this paper may become a platform for modeling and solving assembly line problems with all kind of side constraints, which are typical for real-world industrial applications.

## References

1. A. Aggoun and N. Beldiceanu. Extending CHIP in order to solve complex scheduling and placement problems. *Mathl. Comput. Modelling*, 17(7):57 – 73, 1993.
2. R. K. Ahuja, T. L. Magnanti, and J. B. Orlin. *Network flows : theory, algorithms and applications*. Prentice Hall, 1993.
3. N. Beldiceanu, H. Simonis, Ph. Kay, and P. Chan. The CHIP system, 1997. http://www.cosytec.fr/whitepapers/PDF/english/chip3.pdf.
4. A. Bockmayr and T. Kasper. Branch-and-infer: A unifying framework for integer and finite domain constraint programming. *INFORMS J. Computing*, 10(3):287 – 300, 1998.
5. E. A. Boyd. Polyhedral results for the precedence-constrained knapsack problem. *Discrete Appl. Math.*, 41(3):185–201, 1993.
6. C. E. Ferreira, A. Martin, and R. Weismantel. Solving multiple knapsack problems by cutting planes. *SIAM J. Optim.*, 6(3):858–877, 1996.
7. F. Focacci, A. Lodi, and M. Milano. Cutting planes in constraint programming: A hybrid approach. In *Principles and Practice of Constraint Programming, CP'2000, Singapore*, pages 187 – 201. Springer, LNCS 1894, 2000.
8. E. S. Gottlieb and M. R. Rao. The generalized assignment problem: Valid inequalities and facets. *Math. Program., Ser. A*, 46(1):31–52, 1990.
9. S. Heipcke. *Combined modelling and problem solving in mathematical programming and constraint programming*. PhD thesis, Univ. Buckingham, 1999.
10. J. N. Hooker, G. Ottosson, E. S. Thorsteinsson, and H.-J. Kim. On integrating constraint propagation and linear programming for combinatorial optimization. In *Sixteenth National Conference on Artificial Intelligence, AAAI'99*, 1999.
11. B. Korte and J. Vygen. *Combinatorial Optimization: Theory and Algorithms*. Springer, 2000.
12. M. W. Padberg. (1,k)-configurations and facets for packing problems. *Math. Program.*, 18:94–99, 1980.
13. K. Park and S. Park. Lifting cover inequalities for the precedence-constrained knapsack problem. *Discrete Appl. Math.*, 72(3):219–241, 1997.
14. P. Refalo. Linear formulation of constraint programming models and hybrid solvers. In *Principles and Practice of Constraint Programming, CP'2000, Singapore*, pages 369 – 383. Springer, LNCS 1894, 2000.
15. R. Rodosek, M. G. Wallace, and M. T. Hajian. A new approach to integrating mixed integer programming and constraint logic programming. *Annals of Operations Research*, 86:63 – 87, 1999.
16. A. Scholl. *Balancing and sequencing of assembly lines, 2., rev. ed.* Physica-Verl., 1999.
17. R. L. M. J. van de Leensel, C. P. M. van Hoesel, and J. J. van de Klundert. Lifting valid inequalities for the precedence constrained knapsack problem. *Math. Program.*, 86(1):161– 186, 1999.
18. L. Wolsey. *Integer programming*. Wiley, 1998.